\pdfoutput=1
\documentclass[pra,twocolumn,superscriptaddress,showpacs,floatfix]{revtex4}

\usepackage{epsfig}
\usepackage{epsf}
\usepackage{bm}                 
\usepackage{amsmath}
\usepackage{amssymb}
\usepackage{dcolumn}
\usepackage{graphicx}
\usepackage{psfrag}
\usepackage{latexsym}
\graphicspath{{./}{figs/}}	  

\newcommand{\beq}{\begin{equation}}
\newcommand{\eeq}{\end{equation}}
\newcommand{\beqa}{\begin{eqnarray}}
\newcommand{\eeqa}{\end{eqnarray}}
\newcommand{\ket} [1] {\vert #1 \rangle}
\newcommand{\bra} [1] {\langle #1 \vert}

\newcommand{\proj}[1]{\ket{#1}\bra{#1}}

\begin{document}

\title{ Wigner tomography of two qubit states and quantum cryptography.}

\author{Thomas Durt*, Alexander Ling**,\\ Antia Lamas-Linares***, Christian Kurtsiefer***\\{\footnotesize\it *TENA VUB 
Pleinlaan 2 1050 Brussels Belgium \& thomdurt@vub.ac.be}\\{\footnotesize\it **Centre for Quantum Technologies and Temasek Labs, }\\{\footnotesize\it National University of Singapore, 117543 Singapore,  }\\{\footnotesize\it \& phylej@nus.edu.sg }
\\{\footnotesize\it ***Centre for Quantum Technologies and Department of Physics, }\\{\footnotesize\it National University of Singapore, 117543 Singapore, }\\{\footnotesize\it \& antia@quantumlah.org, christian.kurtsiefer@gmail.com }}

\maketitle
{\it Abstract: Tomography of the two qubit density matrix shared by Alice and Bob is 
an essential ingredient for guaranteeing an acceptable margin of confidentiality during the 
establishment of a secure fresh key through the Quantum Key Distribution (QKD) scheme. We show how the Singapore protocol for key
 distribution is optimal from this point of view, due to the fact that it is based on so called 
 SIC POVM qubit tomography which allows 
 the most accurate full tomographic reconstruction of an unknown density matrix on the basis of a restricted set of experimental data. We illustrate with the help of experimental data the deep connections that exist between SIC POVM tomography and discrete Wigner representations. We also emphasise 
 the special role played by Bell states in this approach 
 and propose a new protocol for Quantum Key Distribution during which a third party is 
 able to concede or to deny A POSTERIORI to the authorized users the ability to build a fresh cryptographic key.
 } 

\section{Introduction}
The goal of quantum cryptographic protocols is to  distill a fresh cryptographic key 
from data encoded in non-commuting bases. In order to guarantee the confidentiality of the 
key, it is essential that Alice and Bob, the authorized users of the channel, check the correlations between their respective signals 
in order to estimate the noise present along the transmission line. This procedure imposes severe constraints to an hypothetical 
eavesdropper (Eve) who cannot manipulate the signal at will and sees her freedom of action seriously limited by the control performed by Alice and Bob. 
This control procedure is optimal when the correlations tested by Alice and Bob are such that
 they allow them to carry out a full tomography of the signal \cite{tomo1}.
  This the case for instance with the so called 6 states protocol \cite{bruss} during which Alice and Bob analyze their 
qubits in three mutually unbiased bases, which allows them to reconstruct the full density matrix of the signal. Due to the fact that 
in realistic situations the length of the key is always finite and that it is preferable that Alice and Bob do not sacrifice too 
many data during their check of the correlations, the optimal tomographic 
procedure is the one for which Alice and Bob are able to maximize the accuracy of their estimation of the signal, keeping fixed the quantity of data that they sacrifice in order to estimate the correlations. This problem has been studied in the past and it appears that the optimal 
procedure for performing full tomography (according to this figure of merit) is to measure a 
Symmetric Informationally Complete Positive Operator Valued Measure (SIC POVM) \cite{caves,Alla04,JRe04}. This approach is at the core of the
 so called Singapore protocol \cite{tomo2singap}
  for Quantum Key Distribution where the signal is encrypted in such a way that Alice and Bob 
 directly measure the SIC POVM distribution of their respective bits,
  and are consequently able to reconstitute the two qubit density matrix of the full signal.
 
 In the present paper we shall discuss certain advantages of the double qubit SIC POVM
  scheme from the point of view of tomography (section \ref{sec2}). We shall then establish a relation between SIC POVM
   tomography and the discrete Wigner representation  (section \ref{sec3} and appendix).
    We shall also show (section \ref{sec4}) that the Bell states constitute 
  the optimal signal for the establishment of a cryptographic key. 
  In the same section, we shall show how
   it is possible thanks to slight modifications of the Singapore protocol \cite{tomo2singap} to provide to
    a third party who controls the source (Charles) the ability
     to deny to Alice and Bob the possibility to build a fresh key although they already recorded all the results of their measurements. Charles has nevertheless the possibility to choose a posteriori to 
 provide them the possibility to do so, by communicating to them the relevant information 
 on a classical communication line. Besides, even in the case that he allows them to establish a fresh key, Charles remains ignorant of the content of this key.

It is worth noting that, due to the fact that it is impossible to amplify a quantum cryptographic key by classical amplifiers, which prohibits the use of standard communication channels, the installation of a communication line for QKD will remain an expensive investment. It is thus probable that in the case that quantum cryptography gets succesfully commercialised, Alice and Bob,
 the authorized users  of a quantum cryptographic channel, will not own it personnally but they will rather rent it to its owner (Charles).  In the case that Alice and Bob
   rent the line to Charles in order to exchange a key, the protocol described in 
   this paper presents obvious advantages regarding the possible commercialisation of 
   Quantum Key Distribution. 
   
   Finally, we shall present in the section \ref{sec5} experimental results that show that within an error of a few percent, all the theoretical concepts developed in this paper are concretely realisable.
\section{About two qubit tomography.\label{sec2}}
\subsection{Optimal PVM tomography.}The estimation of an unknown state is one of the important problems in
quantum information and quantum computation \cite{Na1,N1}. Traditionnally,
 the estimation of the $d^2-1$ parameters that characterize the density matrix of a single 
qu$d$it consists of realising $d+1$ independent von Neumann measurements 
(also called Projection-Valued-Measure measurements or PVM measurements in the litterature) on the system.

 As it was shown in \cite{ivanovic,WoottersF},
    the PVM approach to tomography can be optimised regarding redundancy during the acquisition of the data. Optimality according to this particular figure
     of merit is achieved when
     the $d+1$ bases in which the PVM measurements are performed are ``maximally independent'' or ``minimally overlapping'' so to say when they are mutually unbiased (two
     orthonormal bases of a $d$ dimensional Hilbert space are said to be mutually 
  unbiased bases (MUB's) if whenever we choose one state in the first basis, and a second state in the second basis,
   the modulus squared of their in-product is equal to $1/d$) \cite{ivanovic,WoottersF,JSh60}.  It is well-known that,
    when the dimension of the Hilbert space is a prime power, there exists a 
   set of $d+1$ mutually unbiased bases \cite{ivanovic,WoottersF,india}.  This is the case for instance with the bases that
    diagonalize the generalised Pauli operators \cite{india,TD05}. Those unitary operators form a group which is
     a discrete counterpart of the Heisenberg-Weyl group, the group of displacement operators \cite{milburn}, that present numerous
      applications in quantum optics and in signal theory \cite{vourdas}.

 For instance, when the system is a spin 1/2 particle, three successive Stern-Gerlach measurements performed along orthogonal directions 
 make
  it possible to infer the values of the 3 Bloch parameters $p_{x},p_{y},$%
and $p_{z}$ defined by
\begin{equation}
\left\{ 
\begin{array}{l}
\left\langle \sigma _{x}\right\rangle =p_{x}=\gamma \sin \theta \cos
\varphi \\ 
\left\langle \sigma _{y}\right\rangle =p_{y}=\gamma \sin \theta \sin
\varphi \\ 
\left\langle \sigma _{z}\right\rangle =p_{z}=\gamma \cos \theta%
\end{array}%
\right.
\end{equation}
 Once we know the value of these parameters, we are able to determine unambiguously the 
 value of the density matrix, making use of the identity
  \begin{equation}
\rho (\gamma ,\theta ,\varphi )=\frac{1}{2}(I+p_{x}\sigma
_{x}+p_{y}\sigma _{y}+p_{z}\sigma _{z})=\frac{1}{2}(I+\overrightarrow{\gamma }.\overrightarrow{\sigma })
\end{equation}
When the qubit system is not a spin 1/2 particle but  consists of the polarisation of a photon, a similar result can
 be achieved by measuring its degree of polarisation in three independent polarisation bases, for instance with
  polarising beamsplitters, which leads to the Stokes representation of the state of
   polarisation of the (equally prepared) photons. 
  
  Tomography through von Neumann measurements presents an inherent drawback:
   in order to estimate the $d^2-1$ independent parameters of the density matrix, $d+1$ measurements must be realised which means that 
  $d^2+d$ histograms of the counting rate are established, one of them being sacrificed after each of the $d+1$ measurements in order to normalize
   the corresponding probability distribution. From this point of view, the number of counting rates is higher than the number of parameters 
   that characterize the density matrix, which is a form of redundancy, inherent to the tomography through von Neumann measurements.

In the case of tomographic protocols for QKD, Alice and Bob necessarily perform locally
 and independently a tomographic process on their respective qubits because they are
  separated in space and are not able in principle to carry out non-local measurements \cite{TD05,sanchez}.
Of course by comparing the data gathered during local measurements they can 
         reconstruct the full density matrix but this procedure is highly data consuming in the case of von Neumann (PVM) tomography.
          For instance, in the case that the carrier of the key is a qubit it requires them 
          to estimate 36 joint probabilities. In the framework of Quantum Key distribution
           where the number of available data is per se limited and where the protocols
            of reconciliation and privacy amplification are per se highly data
             consuming it is better to find a tomographic procedure
              that minimizes the redundancies in the data acquisition.
               We shall discuss this procedure in the next section.

 \subsection{Optimal qubit POVM's for tomography.} 
  It is known that a more general class of measurements exists that generalises the von Neumann (PVM) measurements. This class is represented by the Positive-Operator-Valued
    Measure (POVM) measurements \cite{QCQ}, of which only a reduced subset,
     the Projection-Valued-Measure (PVM) measurements correspond to the von Neumann measurements. 
  The most general POVM can be achieved by coupling the system A to an ancilla or assistant B and performing a
   von Neumann measurement on the full system. When both the system and its assistant are 
   qu$d$it systems, the full system 
   belongs to a $d^2$ dimensional Hilbert space which makes it possible to measure $d^2$ probabilities during a von Neumann measurement performed on the full system.
    As always, one of the counting rates must be sacrificed in order to normalise the probability distribution so that we are left with $d^2-1$ parameters.
     When the coupling to the assistant and the von Neumann measurement are well-chosen,
      we are able in 
  principle to infer the value of the density matrix of the initial qu$d$it system from the 
  knowledge of those $d^2-1$ parameters, in which case
   the POVM is said to be Informationally Complete (IC). Obviously, this approach is 
   optimal in the sense that it minimizes the number of counting rates (thus of independent 
   detection processes) 
   that must be realised during the tomographic process. In practice, the implementation of this class of optimal POVMs is simple and has advantages of its own compared to the usual polarization measurements based on von Neumann projections \cite{qlah}.

     IC POVM's can also be further optimised regarding the independence of the data collected in different detectors. The so-called covariant 
     Symmetric-Informationally-Complete (SIC) POVM's \cite{caves} provide an elegant solution to this optimisation constraint. A discrete version of the Heisenberg-Weyl group \cite{weyl} also plays an essential 
     role in the derivation of such POVM's which are intimately associated to a set of $d^2$ minimally overlapping projectors
      onto pure qu$d$it states (the modulus squared of their in-product is now equal to $1/\sqrt{d+1}$). 
      
   In the rest of this paper we shall remain exclusively concerned with qubit SIC POVM's. It has been shown in the past, on the basis of different theoretical 
  arguments \cite{caves,JRe04,Alla04}, that the optimal qubit SIC POVM is in one-to-one correspondence with a tetrahedron
   on the Bloch sphere. Intuitively, such tetrahedrons homogenize and minimize the
    informational overlap or redundancy between the four histograms collected during the POVM measurement. 
   Some of such tetrahedrons can be shown to be invariant under the action of the Heisenberg-Weyl group which corresponds
   to so-called Covariant Symmetric Informationnally Complete (SIC) POVM's \cite{caves}. 
   Concretely, during the measurement of such a SIC POVM, four probabilities of firing $P_{00},P_{01},P_{10},P_{11}$ are measured
   which are in one-to-one correspondence with the Bloch parameters $%
p_{x},p_{y},$ and $p_{z}$ as shows the identity 
\begin{equation}
\left\{ 
\begin{array}{l}
P_{00}=\frac{1}{4}\left[ 1+\frac{1}{\sqrt{3}}(p_{x}+p_{y}+p_{z})\right] \\ 
P_{01}=\frac{1}{4}\left[ 1+\frac{1}{\sqrt{3}}(-p_{x}-p_{y}+p_{z})\right] \\ 
P_{10}=\frac{1}{4}\left[ 1+\frac{1}{\sqrt{3}}(p_{x}-p_{y}-p_{z})\right] \\ 
P_{11}=\frac{1}{4}\left[ 1+\frac{1}{\sqrt{3}}(-p_{x}+p_{y}-p_{z})\right]%
\end{array}%
\right.
\label{povm}\end{equation}

  $2\cdot P_{00}$ is the average value of the operator $({1\over 2})(\sigma_{0,0}+
 ({1\over \sqrt 3})(\sigma_{1,0}+\sigma_{0,1}+\sigma_{1,1}))$ (where $\sigma_{i,j}=
\sqrt{(-)^{i\cdot j}} \sum_{k=0}^1(-)^{k\cdot j}\ket{k+i \,\rm{mod.}2}\bra{k} $; actually, $\sigma_{0,0}=Id.$, $\sigma_{0,1}=\sigma_{z}$, $\sigma_{1,0}=\sigma_{x}, \sigma_{1,1}=\sigma_{y}$ ). One can check that this operator is the projector $\ket{\phi}
   \bra{\phi}$ onto the pure state  $\ket{\phi}=\alpha\ket{0}+\beta^*\ket{1}$ with
    $\alpha=\sqrt{1+{1\over \sqrt 3}} $, 
    $\beta^*=e^{{i \pi\over 4}}\sqrt{1-{1\over \sqrt 3}} $. Under the action of the Pauli group it transforms into a
     projector onto one of the four 
    pure states $\sigma_{i,j}\ket{\phi}$; $i,j:0,1$: $\sigma_{i,j}\ket{\phi}\bra{\phi}
    \sigma_{i,j}=({1\over 2})((1-{1\over \sqrt 3})\sigma_{0,0}+
 ({1\over \sqrt 3})(\sum_{k,l=0}^1(-)^{i\cdot l-j\cdot k}\sigma_{k,l}))$
The signs $(-)^{i\cdot l-j\cdot k}$ reflect the (anti)commutation properties of the Pauli group. So, the four parameters $P_{ij}$ are the average values of
 projectors onto four pure states that are ``Pauli displaced'' of each other.
The in-product between them is equal, in modulus, to $1/\sqrt 3=1/\sqrt{d+1}$, with $d=2$ which is the signature of a Symmetric
    Informationnally Complete (SIC) POVM \cite{caves}. One can show 
    \cite{caves,JRe04,Alla04} that
   such tetrahedrons minimize the informational redundancy between the four 
   collected histograms due to the fact that their angular opening is maximal. The number of counting rates necessary in order to realize a tomographic process by a factorisable 
        POVM measurement is optimal and equal to 16 in the two qubit case. Moreover, 
           the double qubit SIC POVM tomographic scheme is optimal among the factorisable two qubit schemes if we consider
            as a figure of merit*\footnote{Let us consider as an illustrating example that conventional, PVM, qubit tomography is realized in 3 nearly linearly dependent bases, very close to each other. This is obviously a bad tomographic process regarding the redundancy of the collected data. In such a situation, certain coefficients of the density matrix must be estimated on the basis of the differences between frequencies collected in the different bases. Those differences are very small parameters when those bases are strongly overlapping, which corresponds to a high value of the determinant $D$ of the matrix that maps the average probabilities of firing that are measured during the experiment onto the coefficients of the density matrix. In such a case the small experimental discrepancies that affect the measured mean frequencies are strongly amplified during the estimation process of the density matrix and a precise tomography requires, in accordance with the law of large numbers, a high number of data. As we see, a ''good'' tomographic process corresponds to a small value of the determinant $D$.} the determinant $D$ of the matrix that 
           maps the joint probabilities of firing that are collected during the experiment onto the coefficients of the density matrix \cite{Alla04}. This determinant is optimal (minimal) for the SIC POVM (tetrahedron process) in the single qubit case and factorizes when the tomographic process does, so that the determinant of the double tetrahedron process is extremal among the determinants of all factorisable tomographic processes. 

 For all these reasons, the SIC POVM approach is at the core of the
 so called Singapore protocol \cite{tomo2singap}
  for Quantum Key Distribution where the signal is encrypted in such a way that Alice and Bob 
 directly measure the SIC POVM distribution of their respective bits.

\section{Qubit SIC POVM's and discrete Wigner distribution.\label{sec3}} 
   \subsection{The single Qubit case.} 

The qubit covariant SIC POVM possesses another 
 very appealing property \cite{Wootters2} which is
    also true in the qutrit case but not in dimensions strictly higher than 3  \cite{gross}:
    the qubit Covariant SIC POVM is 
a direct realisation (up to an additive and a global normalisation constants) of the 
qubit Wigner distribution of the unknown qubit $a$. Indeed, this distribution $W$ is the 
symplectic
 Fourier transform of the Weyl distribution $w$ \cite{Wootters87,durtwig} (defined by the relation 
 $w_{i,j}=(1/2)Tr.(\rho.\sigma_{i,j})$) which is,
   in the qubit case, equivalent (up to a relabelling of the indices) to its
    double qubit-Hadamard or double qubit-Fourier transform:
  
  \beqa W_{k,l}=(1/2)\sum_{i,j=0}^1(-)^{i\cdot l-j\cdot k}w_{i,j}\nonumber\\
  =((1/\sqrt 2)\sum_{i=0}^1(-)^{i\cdot l})((1/\sqrt 2)\sum_{j=0}^1(-)^{-j\cdot k})w_{i,j}.\label{wign}\eeqa

  One can check that $P_{k,l}=(1/\sqrt 3)W_{k,l}+(1-1/\sqrt 3)/4.$ The discrete qubit Wigner distribution directly generalises its
         continuous counterpart \cite{wigner} in the sense that it provides information about the localisation of the qubit
     system in a discrete 2 times 2 phase space \cite{Wootters87,discretewigner2}.
      For instance the Wigner distribution of
      the first state of the computational basis (spin up along $Z$) is equal to 
      $W_{k,l}(\ket{0})=(1/2)\delta_{k,0}$, which corresponds to a state located 
      in the ``position'' spin up (along $Z$), and homogeneously spread in 
      ``impulsion'' (in spin along $X$), in accordance with uncertainty relations \cite{Rubin1}. 
      Similarly, the Wigner distribution of
      the first state of the complementary basis (spin up along $X$) is equal to 
      $W_{k,l}((1/\sqrt 2)(\ket{0}+\ket{1}))=(1/2)\delta_{l,0}$. For information,
       the fidelities that were achieved in recent experimental 
                 realisations of one qubit SIC POVM tomography 
                 were shown to be of the order of 92 \% in a NMR  realisation  \cite{NMR} 
                  and 99 \% in a quantum optical realisation \cite{qlah}.

\subsection{The double Qubit case.\label{blah}} 
        Certain Wigner distributions factorise in the sense that in the
         two qubit case it is possible to measure a two-qubit or quartit Wigner 
  distributions \cite{discretewigner,durtarxive} by measuring simultaneously local qubit 
  SIC POVMs. 
                      For instance a factorisable Wigner distribution derived in 
                      Refs.\cite{discretewigner,durtarxive}
                       in the case $d=4$ is obtained by performing the tetrahedron measurement 
                       on the first qubit and the anti-tetrahedron measurement 
                       on the second qubit. 
                      The tops of the anti-tetrahedron are obtained from the tops of the tetrahedron by performing
                       on the Bloch sphere a central symmetry around the origin. This transformation is not unitary, but we shall now show that the anti-tetrahedron 
                       is equivalent to the tetrahedron, up to a well-chosen unitary transformation, provided we modify the order
                        of its branches 
                       accordingly.
                       
                       The tops of the anti-tetrahedron are obtained from the tops of the tetrahedron by performing
                       on the Bloch sphere a central symmetry around the origin. This transformation is not unitary, but we shall now show
                        that the anti-tetrahedron 
                       is equivalent to the tetrahedron, up to a well-chosen unitary transformation, provided we modify the order
                        of its branches 
                       accordingly. Indeed, as we noted before (Eqn.(\ref{povm})) the probabilities $P_{ij}$ ($i,j=0,1$) of firing of the detectors associated to the tops of the tetrahedron are equal, up to a normalisation factor $\frac{1}{2}$ to the Born probabilities 
                 of transition onto pure states that are ''Pauli displaced'' of each other. For instance 
                 $\left[ 1+\frac{1}{\sqrt{3}}(\sigma_{x}+\sigma_{y}+\sigma_{z})\right]$ corresponds to $P_{00}$ and is the projector onto the pure state  $\ket{\phi}=\alpha\ket{0}+\beta^*\ket{1}$ with
    $\alpha=\sqrt{1+{1\over \sqrt 3}} $, 
    $\beta^*=e^{{i \pi\over 4}}\sqrt{1-{1\over \sqrt 3}} $ that is represented on the Bloch sphere by the vector 
    $\frac{1}{\sqrt{3}}(1,1,1)$. The three other probabilities of firing that 
    are measured during the qubit SIC POVM tomographic process are the Born probabilities of transition to pure states of which the representation on the Bloch sphere can be obtained
     by rotating the representation of $\ket{\phi}$ of 180 degrees around the axes $X$, $Y$ and $Z$. An anti-tetrahedron can
      be obtained from the tetrahedron by complex conjugating the four tops of the tetrahedron. In what follows,
       the $\sigma_{x}$ and $\sigma_{z}$ operators are assumed to contain real coefficients, and $\sigma_{y}$ to be purely 
       complex. On the Bloch sphere, the four tops of the anti-tetrahedron are then easily seen to be
        obtained by performing onto the tops of the tetrahedron a central symmetry through the origin.

The union of the 4 tops of the tetrahedron and of the anti-tetrahedron forms a set of 8 tops which spans a perfect cube of which the 3 axes of symmetry coincide with the $X$, $Y$ and $Z$ axes. 

Now, if we
         rotate the tops of the anti-tetrahedron (tetrahedron) according to the orthogonal 
        transformation $O$ described by the matrix:
        
          \begin{equation}O=\left(\begin{array}{ccc}
0 & 0 & -1\\
0 & -1 & 0\\
-1 &0& 0
\end{array}\right), \label{ooo}\end{equation}
  we obtain the original tetrahedron (anti-tetrahedron). In the following we shall label the tops of the anti-tetrahedrons
 with the same label as for their image through the central symmetry through the origin (so that the labels ($00$), ($01$), ($10$) and ($11$) respectively correspond to the directions 
  $\frac{1}{\sqrt{3}}(-1,-1,-1)$, $\frac{1}{\sqrt{3}}(1,1,-1)$, $\frac{1}{\sqrt{3}}(-1,1,1)$ and $\frac{1}{\sqrt{3}}(1,-1,1)$).

Due to the fact that the tetrahedron is an oriented object in the 3-dimensional space, the images of the 
tops rotated by $O$ and their images obtained after inversion through the origin ($S_{0}=-Id.$) are not necessarily two by two equal: 
 \begin{equation}O.(1,1,1)=(-1,-1,-1)=S_{0}.(1,1,1);\label{equation7}\end{equation} 
 \begin{equation}O.(1,-1,-1)=(1,1,-1)=S_{0}.(-1,-1,1);\label{equation8}\end{equation} 
  \begin{equation}O.(-1,1,-1)=(1,-1,1)=S_{0}.(-1,1,-1)\label{equation9}\end{equation}  and 
  \begin{equation}O.(-1,-1,1)=(-1,1,1)=
S_{0}.(1,-1,-1).\label{equation10}\end{equation} Roughly speaking this means that in order to pass from the tetrahedron-tetrahedron (T-T) configuration
 to the tetrahedron-anti-tetrahedron (T-A) configuration (a very general definition of the precise 
 meaning of what we mean by these words is given in the next section, here their 
 acceptance ought to be made clear by the context), it is sufficient to 
permute a pair of (well-chosen) indices (here (01) and (10)), and to perform a (well-chosen) rotation which corresponds at the level of Bob's qubit Hilbert space to a (well-chosen) unitary transformation.  
                       
                       Henceforth we shall most often in the following refer to the simultaneous measurement 
                       of local SIC POVMs under the label 
                       ``double tetrahedron'' measurement without precising whether we group the 16 joint
                        probabilities assigned to the 8 (4+4) detectors according to the Tetrahedron-Tetrahedron (TT) or Tetrahedron-Antitetrahedron (TA) configuration.
  
\begin{figure}
\begin{center}
\includegraphics[scale=0.4]{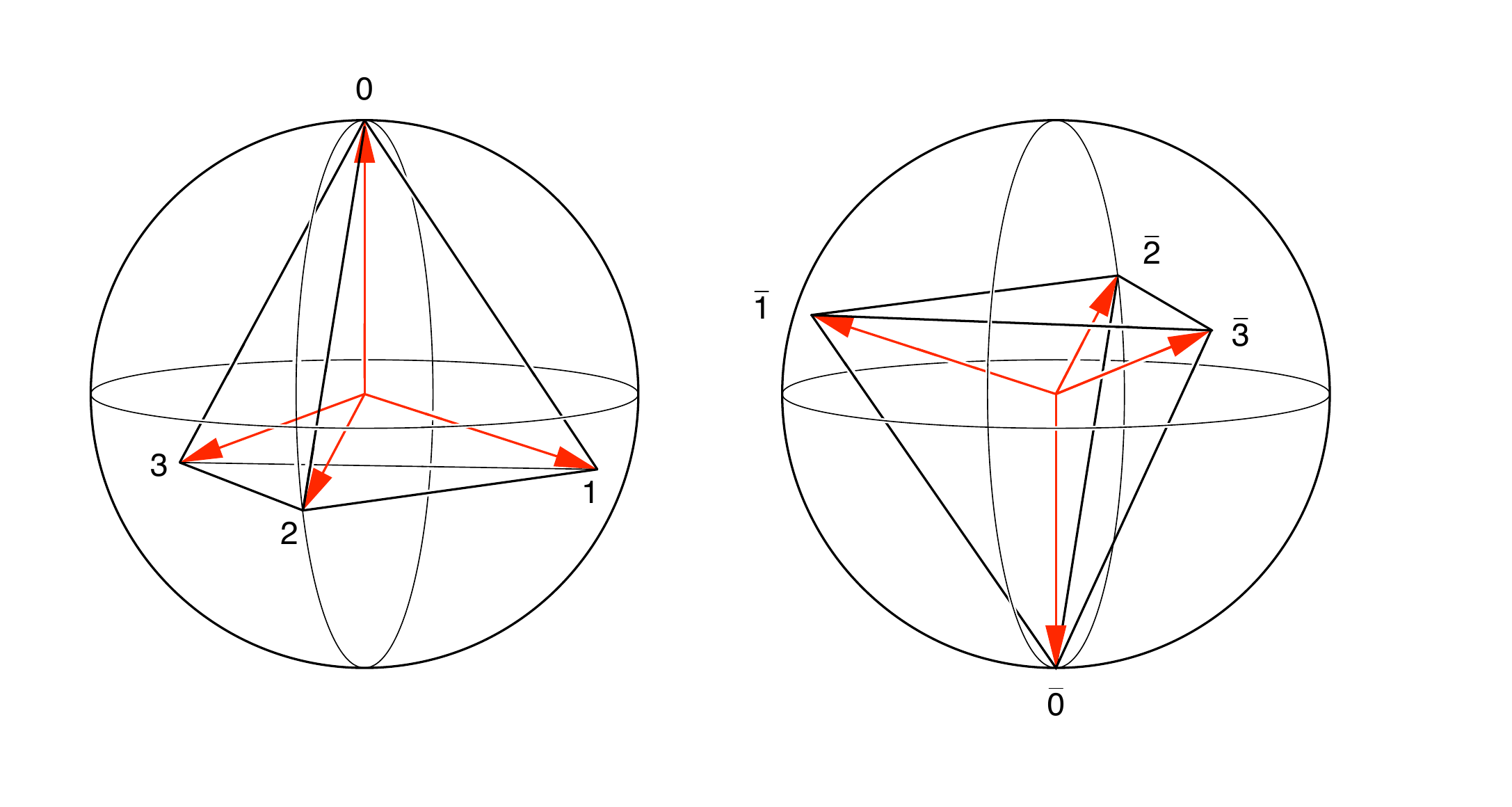}
\end{center}
\caption{Relative orientation of the tetrahedron (left) and antitetrahedron (right) POVMs on the Bloch sphere. Transforming from the tetrahedron POVM to the antitetrahedron POVM involves both a rotation and a relabeling of the possible outcomes.} \label{fig:tetraconcept}
\end{figure}                                         
              \subsection{Double tetrahedron measurement and tomography.}
                
               The relation between the statistics of the double tetrahedron measurement and
                the double Wigner distribution 
               is a straightforward generalisation of the corresponding relation in the 
               single qubit case. Let us denote $P_{k,l}$  (where $k$ and $l$ run from 0 to 3) the 
               joint probability of firing of the $k$th ($l$th) tetrahedron detector on qubit $a$
                (tetrahedron detector on qubit$b$), and let us define $P^{a}_{k}$ ($P^{b}_{l}$) by $P^{a}_{k}= 
               \sum_{l=0}^3P^{ab}_{k,l}$ ($P^{b}_{l}= \sum_{k=0}^3P^{ab}_{k,l}$), then the double 
               Wigner coefficients $W_{k,l}$ can be derived from the statistics of joint detections via the relation
               
               \beq \label{WP}W^{ab}_{k,l}=3\cdot P^{ab}_{k,l}+\sqrt 3\cdot (1-\sqrt 3)/4\cdot  (P^{a}_{k}+P^{b}_{l})+((1-\sqrt 3)/4)^2\eeq
                where the indices $k$ and $l$ run from 0 to 3.

                As the Wigner operators form a basis of the 4 times 4 linear operators and are 
                orthogonal relatively to the Trace-norm, we can reconstruct the full density matrix
                 once we know its double Wigner coefficients, and those are in one to one correspondence with the joint probabilities of firing of the detectors associated to branches of the local ($a$ and $b$) tetrahedrons.
               
                In a next section we shall study the properties of Wigner distributions of Bell
                 states and compare theoretical predictions with data
                 obtained from direct experimental measurement of the correlations.

               \section{Optimal entangled states and cryptography.\label{sec4}}
               \subsection{Optimal entangled states for cryptography.}
               We learnt from the entanglement-based protocol proposed by Artur Ekert \cite{ekert} that it can be useful to exploit non-local correlations between 
               entangled distant quantum systems in order to build a fresh cryptographic key. At this level we did not mention yet explicitly in which bipartite (two qubit) 
               state the signal was prepared. We shall now show that when the key is established by the double tetrahedron protocol (as in Singapore's protocol 
               \cite{tomo2singap}), the optimal strategy is to prepare the pairs of qubits sent to Alice
                and Bob along Bell 
               states  (up to local unitaries). 
               
                The main argument is a symmetry argument. As we mentioned, qubit SIC POVMs are optimal because they are symmetric e.g. such POVMs
                are defined by 
               4 pure states that form an equiangular set (tetrahedron) and are treated on the same footing.
               In order to exploit maximally this symmetry it is natural to try to find entangled states that exhibit either symmetric 
               correlations or symmetric anti-correlations between different branches of the tetrahedrons in the $a$ and $b$ regions. As the tomographic process is full (the Wigner operators form a complete orthonormal basis), once we know the correlation, we also know 
               the state that produces such correlations. We shall show that all states that exhibit symmetric anti-correlations 
               are Bell states up to well-chosen local unitaries (rotations), and that no physically realisable state exhibits symmetric correlations.

 In the case that we consider situations during which the experimental configuration of Alice and Bob's devices is fixed there remains a single degree of freedom that consists of varying the labels of the 4 detectors at each side.
               
               There are then essentially 4!=24 ways (and not $24^2$ ways due to the isotropy of correlations) to group the branches of the tetrahedrons at each side,
                so that we must now investigate the possibility of perfect correlations (anti-correlations) in each of those cases. The 24 permutations between
                 the four tops of the tetrahedron can be realised either by unitary or by anti-unitary transformations so that those permutations 
                 can be partitioned according to their order and their parity (we can define the parity of a permutation in
                  function of the determinant of the 3 x 3 matrix that represents the action at the level of the Bloch sphere of the orthogonal
                   or ``anti-orthogonal'' transformation that realises the corresponding permutation at the level of the tetrahedron branches: even transformations
                    correspond to a value +1 (orthogonal transformations), odd ones to -1 (``anti-orthogonal'' transformations))*\footnote{We could as well have defined the parity of permutations following the mathematical tradition in which the parity of a permutation is defined to be equal to the parity of the number of elementary (two by two) permutations necessary for realizing this permutation.}.

The identity is even and of order 1, there are 
                 3 $(4\cdot 3/2\cdot 2)$ even permutations of order 2 with no fixed top and 6 $(4\cdot 3/2)$ odd permutations of order 2 with two fixed tops,
                  8 $(4\cdot 2)$ even permutations of order 3 with one fixed top
                 and  6 odd permutations of order 4 with no fixed top. These 12 (identity+11) even permutations
                  can be realised by unitary transformations: the 
                 3 even permutations of order 2 with no fixed top are associated to the Pauli $\sigma$ 
                 operators (rotations of 180 degrees around $X$, $Y$ and $Z$) while the 
                 8 even permutations of order 3 with one fixed top correspond to rotations of +/-120 degrees around the axis of the fixed top. 

The 12 remaining (odd) permutations can be
                   decomposed into an odd permutation and
                    an even permutation that can be realised by one among the 12 aforementioned unitary transformations\footnote{Actually, the 12 remaining (odd) permutations can be
                   decomposed into the permutation that fixes $(00)^b$ and $(11)^b$ and permutes $(01)^b$ and $(10)^b$ (see (\ref{equation7},\ref{equation8},\ref{equation9},\ref{equation10})), and
                    an even permutation that can be realised by a unitary transformation (either a 180 degrees rotation
                     around one of the 
                  three main axes $X,Y,Z$ or a 120 degrees rotation around one of the 
                  four branches of the tetrahedron or the identity).}. The reasoning goes as follows: once we impose symmetric correlations (anti-correlations) the Wigner distribution is fully determined and so is the density matrix of the corresponding state. In principle we ought to construct case by case 24 candidate-states for the symmetric correlations and 24 other ones for the symmetric anti-correlations, and check whether the candidates that we find so are physical states (positive definite hermitian operators of trace 1). Due to the unitary equivalence between all even configurations and all odd ones, it is enough to check 2 candidates in the case of perfect correlations and two candidates in the case of perfect anti-correlations, what we shall do now.

               Let us firstly consider a TT (even) configuration and let us assume the existence
                of perfect correlations between {\bf equally labelled} detectors at Alice and Bob's sides so that the joint probabilities obey $P_{k,l}=(1/4)\delta_{k,l};k,l:0,1,2,3$. The expression (\ref{WP}) considerably simplifies in case of symmetric correlations, when all detectors fire with equal probability one fourth at each side. Then we have 
                
\beq W^{ab}_{k,l}=3\cdot P^{ab}_{k,l}-(1/8).\label{idit}\eeq

Making use of this relation we get that $W_{k,l}=(-1/8)+(3/4)\delta_{k,l};k,l:0,1,2,3$. Making use of the identity 
                $\rho=\sum_{k_{a},k_{b},l_{a},l_{b}=0}^1(1/4)W_{k_{a},l_{a}}W_{k_{b},l_{b}}$ where the local
                 (qubit) Wigner
                 operators were defined in Eqn.(
                \ref{wign}), we find by a straightforward computation that 
                
                $\rho$=$(1/4)\cdot (Id.^{ab}+3(\sigma^{a}_X\sigma^{b}_{X}+\sigma^{a}_Y\sigma^{b}_{Y}+
                \sigma^{a}_Z
                \sigma^{b}_{Z}))$=$ \proj{00}+\proj{11}-(1/2) \proj{01}-(1/2) \proj{10}+(3/2) \ket{01}\bra{10}+(3/2) \ket{10}\bra{01}$.
                Such an operator possesses a negative eigenvalue so that it cannot be realised physically.

                If now we impose perfect (and isotropically distributed) anti-correlations then the joint probabilities must obey
                 $P_{k,l}=(1/12)-(1/12)\delta_{k,l};k,l:0,1,2,3$.
                
                By a similar treatment, we find that 

\beq W_{k,l}=(1/8)-(1/4)\delta_{k,l};k,l:0,1,2,3 \label{wigdelta}\eeq

and
                
                $\rho$=$(1/4)(Id.^{ab}-\sigma^{a}_X\sigma^{b}_{X}-\sigma^{a}_Y\sigma^{b}_{Y}-\sigma^{a}_Z
                 \sigma^{b}_{Z})$= $\proj{\Psi^{-}}$ where $\ket{\Psi^{-}}={1 \over \sqrt 2}(\ket{1}^a\ket{0}^b-\ket{0}^a\ket{1}^b)$,
                which is nothing else than the projector onto the singlet state.
        In the previous treatment we assumed that a TT configuration had been chosen. If instead we impose
                 perfect correlations (anti-correlations) in a TA configuration, that we choose at this level to be such that 
                we can replace $W_{k_{b},l_{b}}$ by a similar operator but with opposite signs in front of the operators 
                $\sigma_{x}$, $\sigma_{y}$ and $\sigma_{z}$ (this is a configuration where instead of orienting their tetrahedrons parallely, Alice and Bob choose an experimental set-up in which they orient
                 them anti-parallely*\footnote{In a next section we shall present experimental results that were collected in this configuration. As we have shown in section \ref{blah} such a configuration is equivalent to the TT configuration, up to a well-chosen rotation and a well-chosen (odd) permutation of a pair of local detectors.}.
                 Imposing now perfect correlations (anti-correlations) between equally 
                labelled detectors in this configuration we find by a direct computation that 
                
                $\rho$=$(1/4)(Id.^{ab}-3(\sigma^{a}_X\sigma^{b}_{X}+
                \sigma^{a}_Y\sigma^{b}_{Y}+\sigma^{a}_Z
                \sigma^{b}_{Z}))$ in the first case and 
                
                $\rho$=$(1/4)(Id.^{ab}+\sigma^{a}_X\sigma^{b}_{X}+\sigma^{a}_Y\sigma^{b}_{Y}+\sigma^{a}_Z
                \sigma^{b}_{Z})$ in the second case.

                Such operators are easily shown to admit negative eigenvalues so that no physically
                 acceptable state would allow us 
                to obtain symmetric correlations or anti-correlations in this particular configuration. Actually this is not so astonishing for what concerns symmetric 
                anti-correlations because the partial transposition of a singlet state provides a non-physical state as is well-known \cite{horode}.

As a consequence of the fact that the 12 TT (even) configurations as well as the 12 TA (odd) ones are unitarily equivalent, we have established the following properties:

{\it A. ''In each of the 12 TT configurations there exists exactly one state that exhibits symmetric or isotropic anti-correlations,
 and this state is equivalent to the singlet state up to a well-defined local unitary transformation.''}
 
 {\it B. ''In each of the12 TA (odd) configurations, it is impossible to find a state that exhibits perfect anti-correlations
                   between Alice and Bob's detectors.''}

 {\it C. ''In each of the12 TA (odd) and 12 TT (even) configurations, it is impossible to find a state that exhibits perfect correlations  between Alice and Bob's detectors.''}

 \subsection{Optimal entangled states and generalised cryptographic protocol.}
It is worth noting that when we locally rotate the singlet state around the axis of one of the tops of the tetrahedron, or around its major axes ($X$, $Y$ and $Z$),
                   we still obtain a state that is equivalent, up to a local unitary transformation, to a singlet state and
                   is thus maximally entangled.

In particular, when we locally
                  rotate the singlet state around the one of the major axes ($X$, $Y$ and $Z$) of the tetrahedron,
                   we still obtain a Bell state. Those transformations can be shown to 
                   form a group, the Pauli group or discrete displacement (Heisenberg-Weyl) group*\footnote{The Pauli group is a subgroup of the Clifford group that also counts 24 elements.
                     There is a close connection between the qubit Clifford group that consists of 24 
                     unitary transformations that map Pauli operators onto themselves 
                     under conjugation \cite{Appleby} and the 
group of permutations considered in the present paper. }.

As the four Bell states can be obtained from the singlet state by letting act locally one of the Pauli displacement ($\sigma$) operators onto the singlet state and that these operators map the tetrahedron onto itself, the four Bell states exhibit 
symmetric and perfect anti-correlations at the level of Alice and Bob's detectors.  
 
 Besides, one can easily check that different Bell states anti-correlate one detector at Alice's side with different detectors at Bob's side and vice versa. This is the main feature that we shall exploit in what follows: when Alice and Bob ignore which Bell state they share, they also ignore which of their detectors are anti-correlated and are unable to establish a key.

Let us now assume that a third party (Charles) controls the source, say a singlet state source (actually a source of an arbitrary Bell state would be equally convenient) 
and has the possibility to rotate at will Bob's qubit of 180 degrees around one of the three
 main axes of Bob's tetrahedron (or to do nothing).  If Charles decides to carry out one of those four rotations (Pauli displacements) at random with equal probability (25 percent) without communicating his choices to Alice and Bob, their signal will 
 obviously be totally uncorrelated. If now Charles decides afterwards to inform them about his respective choices they will be able to reconstruct a confidential key (according to the Singapore protocol \cite{tomo2singap} for instance). Due to the symmetry of the correlations exhibited by the Bell states, Charles will remain totally ignorant of the data measured by Alice and Bob, which guarantees  the confidentiality of their key. 
 
 We see thus that Charles possesses the capacity to deny at will to Alice and Bob the authorization to establish a fresh key, even after they measured all the necessary data. This possibility could lead to interesting applications in realistic 
 quantum cryptographic schemes, for instance in the case that Alice and Bob would 
 rent the cryptographic quantum transmission line to Charles, 
its legitimate owner, who would be supposed to be able to control the source and to produce at will one among the 4 Bell states. 
                
In the next section, we represent experimental confirmations of our theoretical predictions concerning anti-correlations exhibited by Bell states.

        \section{ Experimental tomography of Bell states.\label{sec5}}

\begin{figure}
\begin{center}
\includegraphics[scale=1.]{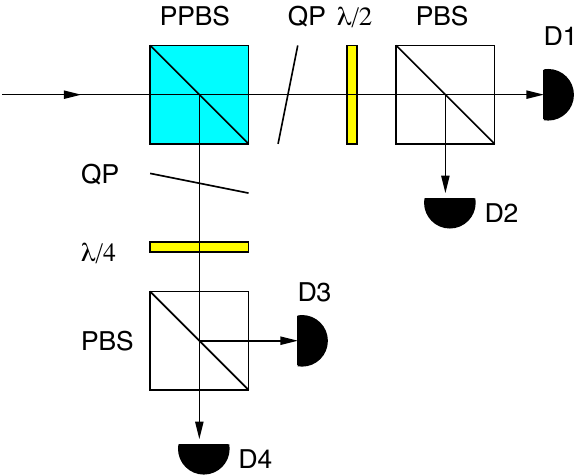}
\end{center}
\caption{Experimental implementation of a single tetrahedron measurement. The incoming photon is incident upon a partially polarizing beam splitter (PPBS) and each of the output arms is modified by the combination of a quartz plate (QP) to modify the phase between H and V polarizarions, and either a quarter wave plate ($\lambda/4$) or a half wave plate ($\lambda/2$) to rotate into the right measurement basis. The final projection is performed by polarizing beam splitters (PBS). The individual photons are detected by Si avalanche photo diodes. Changing from a tetrahedron to an anti-tetrahedron measurement involves only a change in orientation of the waveplates and a relabelling of the detector outputs.}
 \label{fig:setup}
\end{figure}

The afore mentioned Singapore protocol for QKD, in which Alice and Bob share a single state and establish a secret key on the basis of the anti-correlations exhibited by this singlet state when they both realize an optimal SIC POVM onto their respective qubit has been implemented experimentally (see fig. \ref{fig:setup}). Among others this implementation requires a simple polarimetric set-up in order to realize the qubit covariant SIC POVM. This setup was shown in the past to perform tomographic reconstruction of any arbitrary qubit state with high 
fidelity (differing from unity by less than 1 percent) \cite{qlah}. 

\begin{figure}
\begin{center}
\includegraphics[scale=1.]{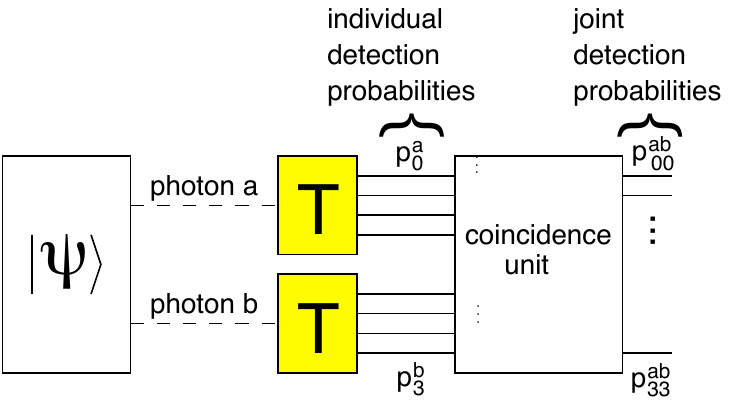}
\end{center}
\caption{Tetrahedron based measurements. Each member of a photon pair is sent to a measurement device implementing either the tetrahedron (T) or anti-tetrahedron (A) POVM (see also Fig.~\ref{fig:setup}). Coincidence timings between a pair of tetrahedron measurements are translated into probabilities by normalizing each joint detection between a particular detector pair to the total number of joint detections.}
 \label{fig:probs-setup}
\end{figure}

We used a pair of such set-ups in order to implement a double tetrahedron measurement on arbitrary Bell states. To prepare the Bell states, we used a Spontaneous Parametric Down Conversion (SPDC) source of entangled photon pairs \cite{paramfluosinga}. The pairs of photons are identified by coincidence timing and the probabilities derived from the raw count rates via normalization to the total number of coincidences (fig.\ref{fig:probs-setup}).

We firstly prepared the singlet Bell state $\ket{\Psi^{-}}={1 \over \sqrt 2}(\ket{1}^a\ket{0}^b-\ket{0}^a\ket{1}^b)$, that we measured in the TT configuration and also in the TA
configuration. 

In a second time, we prepared the Bell state $\ket{\Phi^{+}}={1 \over \sqrt 2}
(\ket{0}^a\ket{0}^b+\ket{1}^a\ket{1}^b)$ that we measured in the
 TT configuration.

In each case we measured the correlation matrices that represent the relative frequency of 
coincidental signals between the detectors of Alice and Bob.

On the basis of those matrices we obtain making use of the relation (\ref{WP}) their respective
 Wigner distributions.

The ideal, theoretical counterpart of the Wigner matrice $W^{TT}(\Psi^{-})$ is expressed by Eqn.(\ref{wigdelta}):

\begin{equation}W^{TT}(\Psi^{-})=(1/8)\left(\begin{array}{cccc}
-1 &1 & 1& 1\\
1 &-1 & 1& 1\\
1 &1& -1& 1\\
1 &1 & 1&-1
\end{array}\right).\label{hamil2}\end{equation}

The counterpart of $W^{TA}_{theo.}(\Psi^{-})$ can be shown to be equal to: 
\begin{equation}W^{TA}_{theo.}(\Psi^{-})= (1/4)\cdot \left(\begin{array}{cccc}
1& 0 & 0 & 0\\
0 & 1&0 & 0\\
0 & 0&1& 0\\
0 & 0 &0 & 1
\end{array}\right). \label{hamil1}\end{equation}

\begin{figure}
\begin{center}
\includegraphics[scale=0.8]{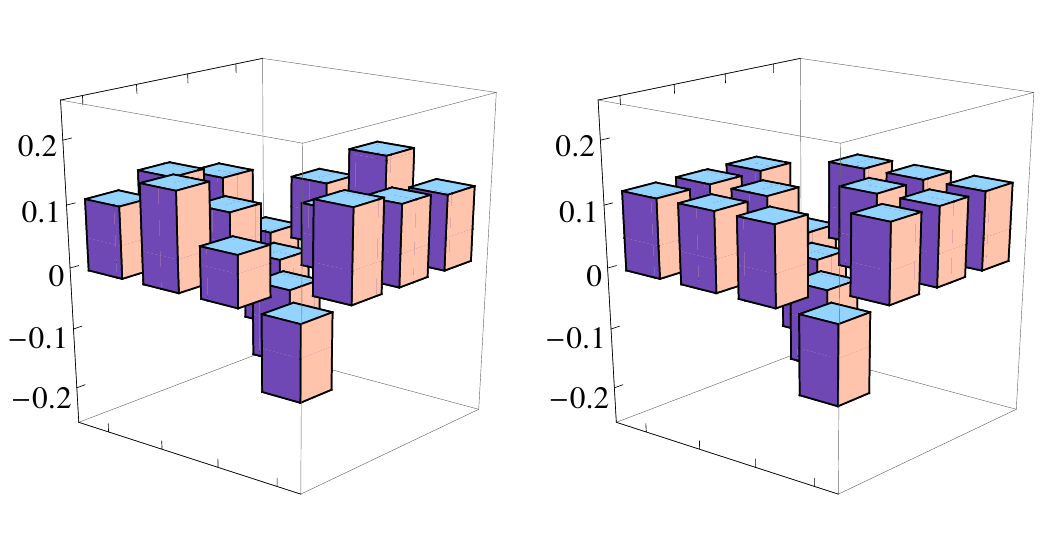}
\end{center}
\caption{Experimental (left) versus theoretical (right, Eqn.(\ref{hamil2})) histograms of the Wigner distribution of the singlet state in the TT configuration.}
 \label{fig:WTT}
\end{figure}

\begin{figure}
\begin{center}
\includegraphics[scale=0.8]{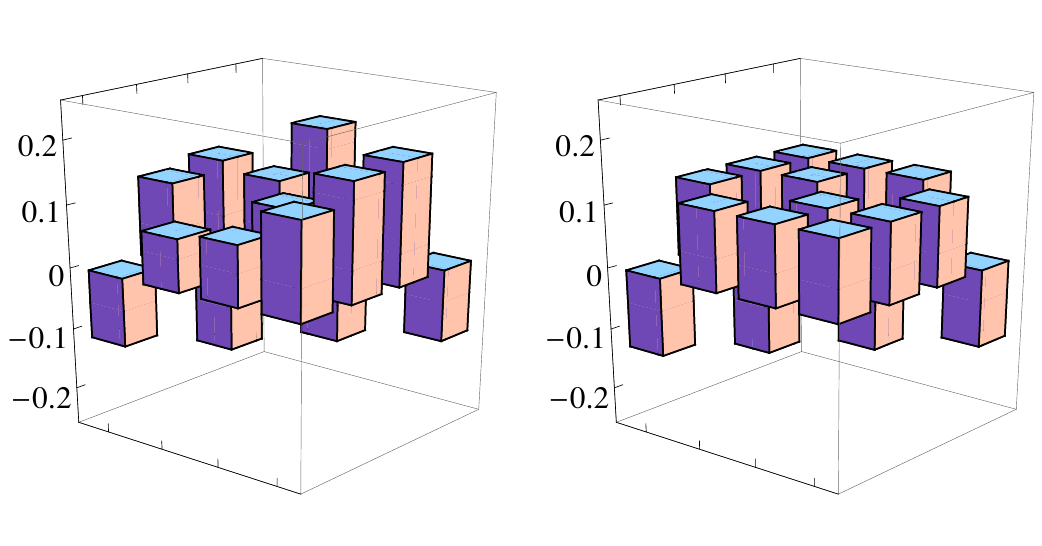}
\end{center}
\caption{Experimental (left) versus theoretical (right, Eqn.(\ref{hamil2})) histograms of the Wigner distribution of the state $\Phi^+$ in the TT configuration.}
 \label{fig:WTT2}
\end{figure}

\begin{figure}
\begin{center}
\includegraphics[scale=0.8]{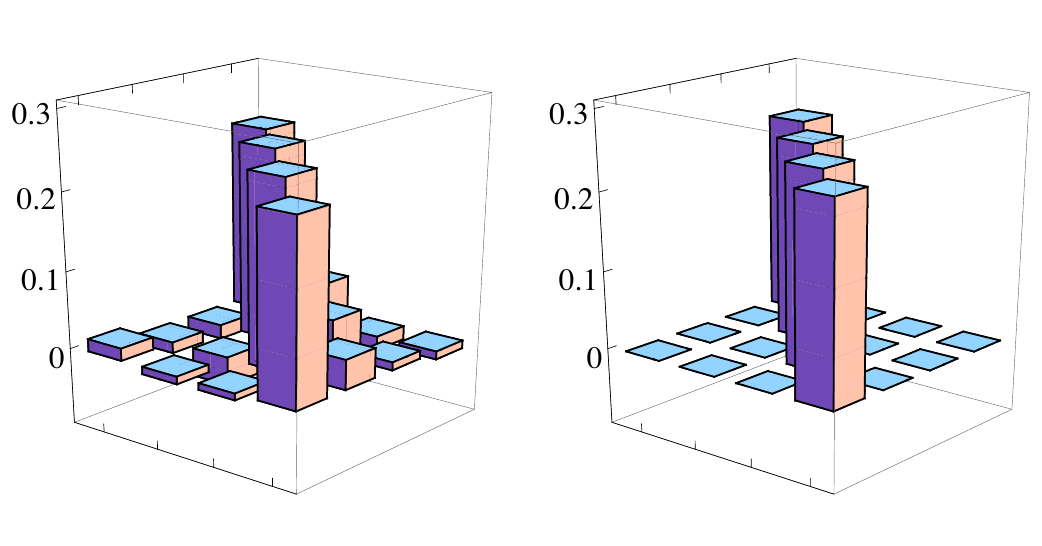}
\end{center}
\caption{Experimental (left) versus theoretical (right, Eqn.(\ref{hamil1})) histograms of the Wigner distribution of the singlet state in the TA configuration.}
 \label{fig:WTA}
\end{figure}

Similarly, one gets that
\begin{equation}W^{TT}(\Phi^{+})=(1/8)\left(\begin{array}{cccc}
1 &1 & 1& -1\\
1 &1 & -1& 1\\
1 &-1& 1& 1\\
-1 &1 & 1&1
\end{array}\right).\label{hamil3}\end{equation}
We controlled that, up to experimental discrepancies and ad hoc reorderings of the coefficients, the experimental Wigner distributions corresponded to their theoretical counterpart. Those results are plotted in figures \ref{fig:WTT} \ref{fig:WTT2} and \ref{fig:WTA}.


We could estimate by a straightforward computation the fidelity of the experimental 
tomographic procedure, making use of the orthonormalisation of the Wigner operators regarding the 
trace norm. The fidelity is equal to four times the sum of the products of the experimentally obtained 
Wigner coefficients with their theoretical counterparts. We obtain so fidelities equal to 
0.960 and 0.956 for the singlet ($\Psi^-$) state in the TT and TA configurations,
 and a bit less for the 
$\Phi^+$ state. The fidelities are less high than for the single qubit tomographic process 
\cite{qlah}, which is not astonishing because in the present case, additive experimental 
errors are likely to occur at the level of the source of entangled states, and also at 
both sides during the local one qubit SIC POVM processes. Moreover supplementary errors could 
also be due to a misalignment between the local tetrahedrons and to finite-size statistical effects (the sample sizes were of the order of $4\cdot 10^4$).

In appendix, the question of the factorisability of two qubit Wigner distributions \cite{durtwig,durtarxive} is discussed in details in relation with the two possible choices of configurations (TT and TA).

 \section{Conclusion.}   
 Protocols for quantum key distribution that allow Alice and Bob to perform full tomography of the signal are optimal for what concerns security against eavesdropping. 
 Due to the fact that in such protocols the amount of data is per se limited, full tomographic protocols based on 
 SIC POVM tomography are also optimal for what concerns the authentification protocol.
 We showed that the natural entangled states that respect the symmetry of SIC POVMs are the Bell states (up to local unitaries). We also showed how their properties make it possible to conceive a protocol during which a third party (Charles) controls the source and 
 is free to concede the authorization to Alice and Bob to establish a key AFTER they measured all the physical data necessary therefore. 
 This possibility opens the way to interesting applications in the case of realistic commercial 
 developments of quantum cryptography; for instance it opens the possibility of delayed on-line payment by the users who rent the cryptographic line.
 
 It is worth mentioning before we end this conclusion that, after all, as was drawn to our attention by a referee, the performances of POVM-based protocols in comparison to those based on PVM's must be relativised. This is due to several reasons. 
 
  Firstly, an apparent advantage of the POVM scheme is that it is not necessary to change constantly and randomly the measurement basis so that we avoid the losses due to basis-reconciliation (sifting) but one should note that the effective gain in bit transfer rate is negligible in comparison to say the BB84 protocol because the POVM-based protocol requires a post-treatment of the correlated data \cite{tomo2singap} that is equally data consuming.    
 
 Secondly one could imagine that even in the (Ekert version of the) BB84 protocol the owner of the line Charles controls the source and encodes the signal at random in the Bell states $\Psi^-$ and $\Phi^{+}$. In such a case Alice and Bob's signal is an incoherent sum with equal weights of perfectly correlated and anti-correlated signals (in the $X$ and $Z$ bases). Such a signal is totally useless for establishing a key before Charles accepts to reveal to Alice and Bob what were his prepared states. The feasibility of a modified BB84 protocol based on entangled photons has been proven experimentally \cite{singaent}.
  
 In principle, the number of copies needed for state estimation can be made arbitrarily smaller than the number of copies needed for the key and it is only when considering finite number effects in a practical implementation that it is advantageous to use a POVM measurement instead of a conventional, PVM, one.

 Finally it should be noted in order not to finish our paper with a sweet and sour note that, beside advantages regarding tomography, POVM emasurements also present effective advantages regarding calibration and stability and remain a promising candidate for quantum key distribution.

\medskip                      
   \leftline{\large \bf Acknowledgment}
\medskip T.D. acknowledges support from the
ICT Impulse Program of the Brussels Capital Region (Project Cryptasc), the IUAP programme of the Belgian government, the grant V-18, the Solvay Institutes for
 Physics and Chemistry, the Fonds voor Wetenschappelijke Onderzoek,
Vlaanderen and last but not least support from the Quantum Lah at N.U.S..

A. L., A. L.-L., and C. K. acknowledge support from ASTAR under SERC grant No. 052 101 0043.

\section{Appendix: Wigner distributions of Bell states.}

\subsection{ Wigner operators and tetrahedrons.}
   Wigner operators or phase space point operators are self-adjoint operators that generalise their continuous counterpart. They are usually defined in such a way that
    (a) their Trace is equal to 1, (b) they
                 are orthogonal with each other and normalised to $d$ (under the Trace norm) (c) if we consider any set 
                 of parallel lines in the phase space, the average of the
                   Wigner operators along one of 
                those lines is equal to a projector onto a pure state, and the averages taken 
                along different parallel lines are projectors onto 
                mutually orthogonal states, (d) the average taken along different and non-parallel lines are projectors
                 onto mutually unbiased states and (e) they are translationally invariant which means that once we know one of them we can find the $d^2-1$ remaining Wigner operators by 
                 letting act the displacement operators onto this one, (f) their sum is equal to $d$ times the identity.

                In a discrete Hilbert space, the phase space is represented by a $d$ times $d$ grid and
                 it is not so simple to define properly the concept of parallelism. 
                Nevertheless, when the dimension is the power of a prime it is possible
                
                (1) to structure a $d$ times $d$ phase space grid in order to 
                find $d+1$ sets of $d$ parallel straight lines with $d$ points such that 
                (parallel) lines from a same set do not intersect, and (non-parallel) lines from
                 different sets intersect in only one point;
                 
                 (2) to find Wigner operators that satisfy the properties (a,b,c,d,e,f) enunciated here 
                 above.
                 
                 The Wigner distribution of a state is a quasi-probability of which the $d^2$ components are
                  equal to the trace of the product of its density matrix with the Wigner operators divided by $d$.

                 In the qubit case, the Wigner operators are defined nearly unambiguously by those properties: as the average of the
                   Wigner operators along lines of different sets are equal to projectors onto
                    pure states from mutually unbiased bases, and that the natural MUBs associated to the Pauli group are
                     the $X$, $Y$ and $Z$ bases, we find that each Wigner operator is equal to the sum of projectors onto 
                    states from those three bases minus the identity operator divided by 2 (in accordance with e.g. equation 
                    (\ref{wign})). Therefore, there exist 8 possible ways to define a qubit Wigner operator, because in each MUB we are free to choose between two basis states. 
                    If moreover, we associate to the 
                    horizontal axis the translation operator $\sigma_{X}$ and to the vertical axis the translation operator 
                    $\sigma_{Z}$, this choice determines all the other Wigner operators (in virtue of translational invariance). There are thus 8 ways to derive a complete set of qubit Wigner operators. It is worth noting that some of them are equivalent up to a Pauli symmetry, and there are four such symmetries, so that there are two sets of  4 Wigner operators that are, inside a given set, unitarily equivalent.
                     We pass from one set to the other through an anti-unitary transformation. In analogy with the section 4.1 where this concept is introduced,
                      we can associate a parity to the Wigner operators, so that 4 of them are 
                      even and four
                      of them are odd; all Wigner distributions then consist of 4 Wigner operators 
                      of same parity that 
                      are translationally equivalent.
                    
                    In order to build a factorisable two qubits Wigner operator we must consider the 64 products of the local
                     $A$ and $B$ Wigner operators and check whether they fulfill the list 
                     of constraints (a) to (f). One can 
                     show \cite{durtwig,durtarxive} that only 
                    32 such products provide an acceptable two-qubits Wigner distribution, and that they consist of products of local qubit operators of different parities (what we denoted the tetrahedron-anti-tetrahedron (T-A) configurations throughout the paper). 
                    
                    It is worth noting that, relatively to the T-T configuration there are few concrete advantages in realising the experiment in the T-A
                configuration from the point of view of tomography. The performances of both 
               tomographic process are very similar and comparable. 
               
               Formally, the interest of performing the T-A configuration is that it provides us a quartit Wigner distribution 
               (as is shown in Ref.\cite{durtwig,durtarxive}, the quartit Wigner distribution does never factorize into a 
               T-T 
               distribution). 
               
               An advantage of having a quartit Wigner distribution at ones disposal is due to the fact that well-chosen marginals of 
                this distribution are proportional to the average values of one-dimensional projectors onto states that belong
                 to MUBs. Those marginals are obtained by summing the Wigner coefficients along straight lines with four
                  elements that belong to the phase-space with 
               16 elements ($k,l$ where $k$ and $l$ run from 0 to 3).  Such a 4 times 4 grid can be structured as
                an affine plane so that it is possible to 
               find 5 sets of four parallel lines with four elements such that parallel lines do not intersect 
               and non-parallel lines intersect in one point only. To each set of four parallel lines (direction)
                corresponds a MUB of which each state corresponds to one line. Three MUBs are factorisable into products of local qubit MUBs, 
                while two MUBs are not factorisable and consist of
                 maximally entangled states \cite{durtwig}. If we want to evaluate the 
               probabilities of transition of an unknown quantum state to those maximally entangled states it is useful to
                perform the tomographic procedure in the T-A configuration because the 
                evaluation of the marginals requires to measure only four joint probabilities, which is less than the 16 joint probabilities that must be measured in order to 
                evaluate the full density matrix. Otherwise, for instance in the context of quantum cryptography it 
                does not really matter wheter we choose the T-T or T-A configuration because all what matters are the physical correlations, which are 
                independent on the relabelling of
                 the branches of the tetrahedron, and we can always pass from one configuration to the other by relabelling 
                 the detectors in an ad hoc manner.
                 
                 \subsection{ Theoretical determination of the Wigner distributions of Bell states.}
                 
                 It is interesting to evaluate the quartit Wigner distributions of Bell states and to compare
                  their expressions with the generic expressions derived in Ref.\cite{paz}. For the singlet state for instance we find that

                  $\rho$=$(1/4)(Id.^{ab}-\sigma^{a}_X\sigma^{b}_{X}-\sigma^{a}_Y\sigma^{b}_{Y}-\sigma^{a}_Z
                 \sigma^{b}_{Z})$

=$(1/4)(W^{tet,a}_{00}W^{antitet,b}_{00}+W^{tet,a}_{01}W^{antitet,b}_{01}
                 +W^{tet,a}_{10}W^{antitet,b}_{10}+W^{tet,a}_{11}W^{antitet,b}_{11})$,
                 
                  where the
                  ``tet'' Wigner operators are defined in accordance with the equation 
                 (\ref{wign}) while the (odd) ``antitet'' operators are obtained from the corresponding (even) ``tet'' operators by inverting the signs of the $\sigma_{X}$, $\sigma_{Y}$, and $\sigma_{Z}$ operators.
                 
                 Now, the phase space grid can be built (among others) by assigning to the  Wigner operator 
                 $W^{tet,a}_{ij}W^{antitet,b}_{kl}$ the coordinates $(m,n)$
                  according to 
                \begin{equation}\label{grid}m=i_{a}\cdot 1+k_{b}\cdot 2, 
                n=j_{a}\cdot 1+l_{b}\cdot 2,\nonumber\end{equation}
                \begin{equation} (m,n:0,1,2,3 \rm{and}\ i_{a},k_{b},j_{a},l_{b}=0,1).\end{equation}
                 
             The Wigner distribution over the phase space grid corresponds thus to the matrix
             
             \begin{equation}W= \left(\begin{array}{cccc}
1/4 & 0 & 0 & 1/4\\
0 & 0&0 & 0\\
0 & 0&0 & 0\\
1/4 & 0 &0 & 1/4
\end{array}\right), \label{hamil1}\end{equation}

via the relation $W_{m,n}=Tr.\rho.W^{a,b}_{m,n}$.

There exist several other ways to define Wigner distributions. For instance we could remain in the T-T configuration where the singlet state 
is associated to the coefficients $W_{mn}=1/8-1/4\delta_{m,n}$ derived at the level of the equation \ref{wigdelta} and keep in mind that we can relabel them in order to 
pass to a T-A configuration by permuting two branches ($01$ and $10$) of say the $b$
 tetrahedron, permuting the roles of $X_{b}$ and $Z_{b}$, and inverting the signs of those axes (due to the rotation of matrix $O$ defined at equation \ref{ooo}). By doing so, we get another Wigner distribution:
 
 \begin{equation}W'=\left(\begin{array}{cccc}
-1/8 &1/8 & 1/8& -1/8\\
1/8 &1/8 & 1/8& 1/8\\
1/8 &1/8 & 1/8& 1/8\\
-1/8 &1/8 & 1/8& -1/8
\end{array}\right), \label{hamil2}\end{equation}

It is shown in Ref.\cite{paz} with the help of very general arguments that the two matrices $W$ and $W'$ do cover all possible
 phase space representations
 of Bell states (up to trivial reorderings associated to translations generated by the local displacement operators). The
  Wigner distributions considered in Ref.\cite{paz} contain our distributions as a special case but, as we see, the reduced set of 32 factorisable
   Wigner distributions \cite{durtwig,durtarxive} considered by us in this paper captures the essential features of the most general case.

\end{document}